\documentclass[11pt]{article}
\usepackage[T1]{fontenc}
\usepackage{lmodern}
\usepackage{amsmath,amssymb,amsthm,mathtools,mathrsfs}
\usepackage{microtype,booktabs,array}
\usepackage[margin=1in]{geometry}
\usepackage{setspace}
\usepackage[authoryear,round]{natbib}
\setcitestyle{authoryear,round,aysep={\ },yysep={;}}
\usepackage[colorlinks=true,linkcolor=black,citecolor=black,urlcolor=blue]{hyperref}
\allowdisplaybreaks
\numberwithin{equation}{section}
\newtheorem{theorem}{Theorem}[section]
\newtheorem{proposition}[theorem]{Proposition}
\newtheorem{corollary}[theorem]{Corollary}
\newtheorem{lemma}[theorem]{Lemma}
\theoremstyle{remark}

\newcommand{\R}{\mathbb R}
\newcommand{\Pp}{\mathbb P}
\newcommand{\E}{\mathbb E}
\newcommand{\pr}{\operatorname{pr}}
\newcommand{\cX}{\mathcal X}
\newcommand{\cO}{\mathcal O}
\newcommand{\cA}{\mathcal A}
\newcommand{\cW}{\mathcal W}
\newcommand{\ind}{\mathbf 1}
\newcommand{\FH}{\mathrm{FH}}
\newcommand{\one}{\mathbf 1}

\title{Sharp Fr\'echet--Hoeffding Bounds under\\Selectively Incomplete Observation}
\author{Rami V. Tabri\footnote{Department of Econometrics and Business Statistics, Monash University. Email: rami.tabri@monash.edu.}}
\date{}
\begin{document}
\maketitle

\begin{abstract}
Incomplete records can sharpen bounds on joint outcome
probabilities even when no pair of outcomes is jointly
observed. I derive sharp pointwise Fr\'echet--Hoeffding
bounds combining known population marginals with
selectively observed records, without restricting the
observation mechanism. I establish an exact reduction
of the identification problem to finite-dimensional
optimization, yielding computable characterizations
of both bounds. Illustrations involving income mobility,
potential outcomes, and split questionnaires demonstrate
how incomplete records can improve identification
relative to the classical bounds.
\end{abstract}


\section{Introduction}

Many questions in empirical economics concern the joint
behavior of outcomes that are not fully observed.
Examples include income persistence across generations,
dependence between potential outcomes, and
multidimensional economic deprivation.
Sample selectivity arises when whether an outcome is observed depends on its value or on other, unobserved outcomes.
Consequently, the distribution of observed records
need not be representative of the population,
complicating identification of the joint outcome
distribution \citep{Heckman1979,Manski1989}.
Even when population marginal distributions are
known from auxiliary information, they do not
determine the dependence among outcomes.
The classical Fr\'echet--Hoeffding (FH) inequalities
\citep{Hoeffding1940,Frechet1951} provide sharp
bounds using these marginals alone.
But incomplete records contain additional
information: they reveal which outcomes are observed
and their realized values.
Can this information sharpen the classical bounds
without imposing restrictions on the observation
mechanism?

The question links sample selectivity to the classical
marginal problem: even under unrestricted observation,
the observed-data law may constrain joint distributions
with known population marginals. The theory of marginal
problems and its connections to Fr\'echet bounds
are well established; see
\citet{Ruschendorf1991}. Copula approaches to
dependence with prescribed marginals are reviewed
by \citet{FanPatton2014}, while
\citet{FanHenry2023} develop vector copulas using
optimal transport. Additional information about dependence can sharpen
these bounds \citep{LuxPap2017,Bartl2022}.
In treatment-effect models, \citet{FanWu2010}
study identification under switching-regime restrictions,
while \citet{FanGuerreZhu2017} characterize the
identifying value of covariate information for
functionals of joint potential-outcome distributions.
Related work develops approaches to partial identification
with incomplete observations using linear programming,
subdistribution extensions, and partially identified
copulas \citep{Zaffalon2002,Torgovitsky2019,FanShiTao2023}.
Combinatorial methods have also been used to study
contingency-table bounds and missing-data compatibility
\citep{DobraFienberg2000,BerrettSamworth2023}.
I consider the identifying content of a different
combination of information: known population marginals
and the complete observed-data law under unrestricted
outcome-dependent observation patterns. The objective
is to characterize the resulting sharp FH bounds
explicitly and determine when incomplete records
improve upon the classical inequalities.

I study a multivariate outcome
$X=(X_1,\ldots,X_d)$ whose coordinates are observed
according to a random subset $S\subseteq\{1,\ldots,d\}$.
For every observation pattern $T$, the observed-data
law identifies the subdistribution
$\mu_T(B)=\Pp(S=T,X_T\in B)$, while the population
marginals $F_j$ are known. I impose no restrictions
on the dependence between $S$ and $X$. The first
result establishes an exact reduction of the
identification problem to a finite allocation problem
for binary threshold indicators.
Every feasible binary allocation can be lifted to a compatible full-data distribution preserving both the original marginal laws and the entire observed subdistributions; equivalently, the mapping from admissible full-data laws to feasible binary allocations is surjective. Consequently, the reduction yields sharp bounds for
arbitrary binary-profile scores, even when the
underlying outcomes are continuous.

I specialize this characterization
to the joint-CDF ordinate
$H(x)=\Pp(X_1\leq x_1,\ldots,X_d\leq x_d)$.
The sharp lower bound allocates threshold failures to
unobserved coordinates, subject to marginal failure
probabilities and eligible pattern masses. This yields
a maximum-flow problem and an explicit minimum-cut
formula \citep{AhujaEtAl1993}. The sharp upper bound
allocates threshold successes to unobserved coordinates,
leading to a fractional packing problem and its
finite-dimensional dual \citep{Schrijver2003}.
Both bounds are attained, with observable necessary
and sufficient conditions for strict FH tightening.
Computationally, the lower endpoint uses network flow,
while the upper endpoint has one variable per occurring
observation pattern and $d$ marginal-resource constraints.

I illustrate the identifying strength of incomplete records in several examples.
In the three-generation income example, the upper bound
on bottom-quintile persistence falls from $0.20$ to
$0.1125$, although no two incomes are jointly observed.
The Supplementary Appendix also treats potential outcomes
and high-dimensional split questionnaires. I focus on population identification, leaving sampling
uncertainty for future research.

Section~\ref{sec:setup} gives the observed-data and
transport setup, Section~\ref{sec:binary} the exact
binary reduction, and Section~\ref{sec:bounds} the sharp
FH bounds. Sections~\ref{sec:discussion} and~\ref{sec:Conclusion}
discuss the main result
and conclude. Proofs are in the Appendix, with further
results and illustrations in the Supplementary Appendix.

\section{Observed-data decomposition and constrained transport}
\label{sec:setup}

Fix an integer $d\geq 2$ and write $[d]=\{1,\ldots,d\}$.
For each $j\in[d]$, let $\cX_j$ be a Polish space with its
Borel $\sigma$-algebra $\mathcal B(\cX_j)$. Set
$\cX=\prod_{j=1}^d\cX_j$, equipped with its product Borel
$\sigma$-algebra $\mathcal B(\cX)$.

Let $(\Omega,\mathcal F,\Pp)$ be a probability space.
The complete outcome vector is the measurable mapping
\[
X=(X_1,\ldots,X_d):
(\Omega,\mathcal F)
\longrightarrow(\cX,\mathcal B(\cX)).
\]
Denote by $H=\Pp\circ X^{-1}$ its joint probability law
and by $F_j=(\pr_j)_\#H$ its $j$th marginal probability
law, where $\pr_j:\cX\to\cX_j$ is the coordinate
projection. For $T\subseteq[d]$, put
$\cX_T=\prod_{j\in T}\cX_j$ and let
$\pr_T:\cX\to\cX_T$ denote the corresponding projection.
For $j\in T$, write $\pr_j^T:\cX_T\to\cX_j$, so that
$\pr_j=\pr_j^T\circ\pr_T$. By convention,
$\cX_\varnothing=\{\ast\}$ is a singleton and
$\pr_\varnothing$ is the constant map. Write
$X_T=\pr_T\circ X$.

Let
\(
S:(\Omega,\mathcal F)
\longrightarrow(2^{[d]},\mathcal P(2^{[d]}))
\)
be a measurable observation pattern: $S(\omega)$
contains exactly those coordinates of $X(\omega)$
that are recorded. The observation mechanism represented by $S$
may arise from nonresponse, survey design,
treatment assignment, or other processes
determining which coordinates are recorded.
No restriction is imposed on the dependence
between $S$ and $X$. The random set $S$
is distinguished from a fixed pattern $T\subseteq[d]$
throughout.

Define the observed-record space and its
$\sigma$-algebra by
\[
\cO=\bigcup_{T\subseteq[d]}(\{T\}\times\cX_T),
\qquad
\mathcal B(\cO)=
\sigma\bigl(
\{\{T\}\times B:
T\subseteq[d],\ B\in\mathcal B(\cX_T)\}
\bigr).
\]
The label $T$ identifies which coordinates
are observed, so records belonging to
different observation patterns are treated
as distinct elements of $\cO$. The observed record is the measurable mapping
$O:\Omega\to\cO$ defined by
\(
O(\omega)=
\bigl(S(\omega),X_{S(\omega)}(\omega)\bigr),
\)
and $P_O=\Pp\circ O^{-1}$ is its law.
For each fixed $T\subseteq[d]$, the observed
subdistribution on $\cX_T$ is
\begin{equation}\label{eq:mu}
\begin{split}
\mu_T(B)
&=\Pp(S=T,X_T\in B),
\quad B\in\mathcal B(\cX_T),\\
p_T
&=\mu_T(\cX_T)=\Pp(S=T),
\qquad
\sum_{T\subseteq[d]}p_T=1.
\end{split}
\end{equation}
Thus $\mu_T$ is a nonnegative finite measure with
total mass $p_T$, not a probability law unless
$p_T=1$. In particular,
$\mu_\varnothing=p_\varnothing\delta_\ast$, and
$P_O(\{T\}\times B)=\mu_T(B)$. The entire collection
$\mu=(\mu_T)_{T\subseteq[d]}$ is identified by the
observed-data law.

For the true full-data law, define the latent
pattern-specific subprobability measures on $\cX$ by
\begin{equation}\label{eq:truepi}
\pi_T^0(B)=\Pp(X\in B,S=T),
\quad B\in\mathcal B(\cX),
\qquad
\pi_T^0(\cX)=p_T.
\end{equation}
When $p_T>0$, $\pi_T^0/p_T$ is the conditional law
of the entire outcome $X$ given $S=T$; when $p_T=0$,
$\pi_T^0$ is the zero measure. Although $\pi_T^0$
is not observed, it satisfies
\[
(\pr_T)_\#\pi_T^0=\mu_T,
\qquad
H=\sum_{T\subseteq[d]}\pi_T^0,
\qquad
F_j=\sum_{T\subseteq[d]}(\pr_j)_\#\pi_T^0
\quad(j\in[d]).
\]

Let $\mathcal M_+(\cX)$ be the finite nonnegative
Borel measures on $\cX$, and let
\(
\mathcal M_p(\cX)=
\{\nu\in\mathcal M_+(\cX):\nu(\cX)=p\}.
\)
For $F=(F_1,\ldots,F_d)$, the admissible family of
candidate latent subprobability measures is
\begin{equation}\label{eq:A}
\begin{split}
\cA(F,\mu)=
\Bigl\{
(\pi_T)_{T\subseteq[d]}
\in\prod_{T\subseteq[d]}\mathcal M_{p_T}(\cX):
\quad&
(\pr_T)_\#\pi_T=\mu_T
\quad(T\subseteq[d]),\\
&
\sum_{T\subseteq[d]}(\pr_j)_\#\pi_T=F_j
\quad(j\in[d])
\Bigr\}.
\end{split}
\end{equation}
Every member specifies a possible full-data law,
not necessarily the true collection $(\pi_T^0)_T$.
Such a law may be realized on the canonical space
$\cX\times2^{[d]}$, and need not be realized on the
original probability space $(\Omega,\mathcal F,\Pp)$.

For subsequent use, define the residual marginal
measure
\begin{equation}\label{eq:residual}
R_j
=
F_j-\sum_{T:j\in T}(\pr_j^T)_\#\mu_T,
\qquad
m_j=R_j(\cX_j)
=\sum_{T:j\notin T}p_T.
\end{equation}
Under the maintained probability model,
\(
R_j(B)=\Pp(j\notin S,X_j\in B)
\geq 0,
\; B\in\mathcal B(\cX_j).
\)
Thus $R_j$ is a nonnegative measure representing
the contribution to the population marginal law
of coordinate $j$ from observations under which
that coordinate is missing.

\begin{proposition}[Nonemptiness and weak compactness]
\label{prop:compact}
Under the probability model of Section~2, the
admissible family $\cA(F,\mu)$ is nonempty and
compact in the product of the weak topologies on
\(
\prod_{T\subseteq[d]}\mathcal M_{p_T}(\cX).
\)
\end{proposition}

For any bounded Borel cost $c:\cX\to\R$, define
the sharp endpoints for $\E[c(X)]$ by
\begin{align}\label{eq:bounds}
L_c
=
\inf_{\pi\in\cA(F,\mu)}
\sum_{T\subseteq[d]}
\int_{\cX}c(z)\,\mathrm d\pi_T(z),\quad
U_c
=
\sup_{\pi\in\cA(F,\mu)}
\sum_{T\subseteq[d]}
\int_{\cX}c(z)\,\mathrm d\pi_T(z).
\end{align}
Convexity of $\cA(F,\mu)$ implies that the identified
set is an interval with closure $[L_c,U_c]$.
By Proposition~\ref{prop:compact}, continuous bounded
costs attain both extrema; compactness alone does not
ensure attainment for bounded Borel costs. For the
threshold costs below, attainment follows from
Theorem~\ref{thm:lift}.

Equation~\eqref{eq:bounds} allocates mass across the
latent measures $(\pi_T)_T$ while reproducing each
recorded-coordinate distribution and all population
marginals. These constraints need not identify the joint law.

The identification problem also has a constrained
multimarginal transport representation: couple an
observed record $o=(T,y_T)$ with a complete outcome
$z\in\cX$, preserving both the observed-record law and
the population marginals, and requiring the completion
to agree with the recorded coordinates. No restriction
on the observation mechanism is imposed.

\begin{proposition}[Sharp constrained-transport representation]
\label{prop:ot}
Under the conditions of Proposition~
\ref{prop:compact}, the sharp endpoints in
\eqref{eq:bounds} equal the infimum and supremum of
\(
\int_{\cO\times\cX}c(z)\,\mathrm d\gamma(o,z)
\)
over probability measures $\gamma$ on
$\cO\times\cX$ satisfying
\begin{equation}\label{eq:otconstraints}
\gamma_O=P_O,\qquad
\gamma_{X_j}=F_j\quad(j\in[d]),
\qquad
\gamma(K)=1,
\end{equation}
where
\(
K=
\bigl\{
((T,y_T),z)\in\cO\times\cX:
\pr_T(z)=y_T
\bigr\}
\)
is the observation-consistency relation.
\end{proposition}

Here $\gamma_O$ and $\gamma_{X_j}$ denote,
respectively, the observed-record marginal
and the $j$th latent-coordinate marginal of
$\gamma$. The condition $\gamma(K)=1$ ensures
that every completed outcome agrees with
its recorded coordinates. The equivalence
follows by pushing each $\pi_T$ forward under
$z\mapsto((T,\pr_T(z)),z)$, and conversely
restricting $\gamma$ to each observation
pattern. 

The transport formulation makes explicit that
the observed-data law imposes restrictions
beyond the population marginals alone. It
also provides a common framework for deriving
sharp bounds and their dual representations.
Related work by \citet{FanPassShi2026} uses
optimal transport to characterize identified
sets in moment models with unlinked datasets.
The formulation here instead accommodates arbitrary
observation patterns and their associated
subdistributions, and focuses on sharp bounds
for joint-CDF probabilities. Since the
joint-CDF indicator is generally discontinuous,
I do not rely on continuous-cost transport
duality. Section~\ref{sec:binary} instead
establishes an exact finite-dimensional
reduction, whose intersection structure yields
the explicit sharp bounds in
Section~\ref{sec:bounds}.

%The transport formulation makes explicit that
%the observed-data law imposes restrictions
%beyond the population marginals alone. It
%also provides a common framework for deriving
%sharp bounds and their dual representations.
%The primary interest is the joint-CDF indicator, which is generally discontinuous.
%Rather than rely on continuous-cost
%transport duality, Section~\ref{sec:binary}
%establishes an exact finite-dimensional
%reduction. The special structure of the
%intersection cost then yields the explicit
%sharp bounds developed in Section~\ref{sec:bounds}.

\section{Exact threshold reduction and sharpness}
\label{sec:binary}

The constrained transport problem of Section~\ref{sec:setup}
is measure-valued. I focus on the joint-CDF ordinate
$H(x)$, which depends only on threshold indicators,
and establish an exact finite-dimensional reduction. For the remainder of the paper, take $\cX_j=\R$ for each
$j\in[d]$. Fix $x=(x_1,\ldots,x_d)\in\R^d$ and define
\begin{equation}\label{eq:D}
\begin{gathered}
B_j^1=(-\infty,x_j],\qquad B_j^0=(x_j,\infty),\\
D_j=\ind\{X_j\in B_j^1\},\qquad
q_j=F_j(B_j^1),\qquad H(x)=\Pp(D=\one_d),
\end{gathered}
\end{equation}
where $D=(D_1,\ldots,D_d)$ and $\one_d=(1,\ldots,1)$.
Thus $q_j$ is the known population probability that $D_j=1$.

For each observation pattern $T\subseteq[d]$ and
$a\in\{0,1\}^T$, let
\(
B_T^a=\prod_{j\in T}B_j^{a_j},\;
D_T=(D_j)_{j\in T}.
\)
The observed-data law identifies the threshold-category
probabilities
\begin{equation}\label{eq:theta}
\theta_{T,a}=\mu_T(B_T^a)=\Pp(S=T,D_T=a),\qquad
\sum_{T\subseteq[d]}\sum_{a\in\{0,1\}^T}\theta_{T,a}=1.
\end{equation}
These probabilities classify observed records by their
observation patterns and the threshold indicators of their
recorded coordinates.

To describe possible joint laws of $(S,D)$, consider the
latent cell masses
\(
w_{T,b}=\Pp(S=T,D=b),\;
T\subseteq[d],\; b\in\{0,1\}^d.
\)
They are not generally identified, because the threshold
indicators of missing coordinates are unobserved. A candidate
allocation $w=(w_{T,b})_{T,b}$ must reproduce both the observed
category probabilities $\theta_{T,a}$ and the population marginal
probabilities $q_j$. This gives the polytope
\begin{equation}\label{eq:W}
\cW(\theta,q)=\Bigl\{w\in\R_+^{4^d}:\;
\sum_{b:b_T=a}w_{T,b}=\theta_{T,a}
\quad(T\subseteq[d],\ a\in\{0,1\}^T),
\sum_{T\subseteq[d]}\sum_{b\in\{0,1\}^d}
b_jw_{T,b}=q_j\quad(j\in[d])\Bigr\}.
\end{equation}
The first constraints reproduce the observed threshold
categories; the second reproduce the known marginal probabilities.

However, the original subdistributions $\mu_T$ specify the
joint distributions of recorded values within each pattern,
not only their threshold categories. It remains to determine whether the polytope
$\cW(\theta,q)$ characterizes exactly the binary
cell allocations generated by compatible full-data
laws satisfying all the measure constraints in
\eqref{eq:A}. The following result establishes
this equivalence.

\begin{theorem}[Exact binary reduction and lifting]
\label{thm:lift}
Under the probability model of Section~\ref{sec:setup}, an array
$w=(w_{T,b})_{T,b}\in\R^{4^d}$ belongs to
$\cW(\theta,q)$ if and only if there exists a family
$(\pi_T)_{T\subseteq[d]}\in\cA(F,\mu)$ such that
\[
\pi_T\left(\prod_{j=1}^d B_j^{b_j}\right)=w_{T,b},
\qquad T\subseteq[d],\quad b\in\{0,1\}^d.
\]
Consequently, for every score $g:\{0,1\}^d\to\R$, the sharp bounds on $\E[g(D)]$ over all compatible
full-data laws are
\begin{align*}
L_g(\theta,q)
=\min_{w\in\cW(\theta,q)}
\sum_{T\subseteq[d]}\sum_{b\in\{0,1\}^d}
g(b)w_{T,b},\quad
U_g(\theta,q)
=\max_{w\in\cW(\theta,q)}
\sum_{T\subseteq[d]}\sum_{b\in\{0,1\}^d}
g(b)w_{T,b}.
\end{align*}
Both extrema are attained.
\end{theorem}

The proof constructs compatible
full-data distributions by combining observed conditional laws
within threshold categories with restrictions of the residual
marginal measures. The construction preserves the entire
subdistributions $\mu_T$ and marginal laws $F_j$, not merely
their threshold probabilities; no continuity of $F_j$ is needed.

More generally, Theorem~\ref{thm:lift} characterizes the
sharp identified set for the \emph{distribution} of $D$.
Writing $p_b=\Pp(D=b)$, the set of admissible vectors
$p=(p_b)_{b\in\{0,1\}^d}$ is precisely
\(
\left\{\left(\sum_{T\subseteq[d]}w_{T,b}\right)_{
 b\in\{0,1\}^d}:w\in\cW(\theta,q)\right\}.
\)
It is a nonempty compact convex polytope, so the theorem
provides sharp bounds and joint identifying restrictions for
any collection of binary-profile scores.

This generality includes rank-transition probabilities
\citep{ChettyEtAl2014}, such as
$\Pp(D_2=0\mid D_1=1)=\E[D_1(1-D_2)]/q_1$ for $q_1>0$,
and counting-based poverty measures
\citep{AlkireFoster2011}, including
$\Pp(\sum_jD_j\geq k)$ and the adjusted headcount
$\E[(\sum_jD_j/d)\ind\{\sum_jD_j\geq k\}]$.
For the intersection score $g_\cap(b)=\prod_{j=1}^db_j$,
Section~\ref{sec:bounds} gives explicit formulas beyond
the general linear programs.

\section{Sharp intersection bounds and identifying information}
\label{sec:bounds}

Theorem~\ref{thm:lift} expresses the sharp bounds for any
binary-profile score as linear programs over
$\cW(\theta,q)$. I now specialize to the intersection
score $g_\cap(b)=\prod_{j=1}^d b_j$, whose expectation
is the joint-CDF ordinate $H(x)$. The particular
structure of this event permits a more explicit solution
than the general $4^d$-variable allocation problem.

For each observation pattern $T\subseteq[d]$,
write $M(T)=[d]\setminus T$ for its missing
coordinates, and let $1_T$ denote the all-one
element of $\{0,1\}^T$. Define the
pattern-specific eligible mass
\(
a_T=\theta_{T,1_T}
=\mu_T\left(\prod_{j\in T}B_j^1\right),
\)
and the total eligible mass
\(
A=\sum_{T\subseteq[d]}a_T.
\)
A record with pattern $T$ can belong to the
intersection event only if all its recorded
indicators equal one. Thus $a_T$ is the
eligible mass within pattern $T$, and $A$
is the total mass eligible for the intersection.
For $T=\emptyset$, eligibility is vacuous
and $a_\emptyset=p_\emptyset$.

Recall the residual measures $R_j$ from
\eqref{eq:residual}. Their threshold masses are
%\begin{equation}\label{eq:r}
$r_j^e=R_j(B_j^e)
=\Pp(j\notin S,D_j=e),
\;
e\in\{0,1\},\; r_j^0+r_j^1=m_j.$
%\end{equation}
Although the missing indicators are not recorded,
these masses are identified: $r_j^1$ equals $q_j$
minus the observed mass with $D_j=1$, and $r_j^0$
equals $1-q_j$ minus the observed mass with
$D_j=0$. Consequently, every
$w\in\cW(\theta,q)$ satisfies
\(
\sum_{T:j\notin T}\ \sum_{b:b_j=e}w_{T,b}
=r_j^e,
\; j\in[d],\; e\in\{0,1\}.
\)
These identities constrain how eligible records
can be completed. To exclude an eligible record
from the intersection, at least one of its missing
indicators must be assigned zero; to include it,
all its missing indicators must be assigned one.
The available masses $r_j^0$ and $r_j^1$ therefore
act as pattern-dependent resources for minimizing
and maximizing the intersection probability,
respectively.

\begin{theorem}[Sharp multivariate Fr\'echet--Hoeffding bounds]
\label{thm:sharp}
Under the conditions of Theorem~\ref{thm:lift}, at every fixed
$x\in\R^d$, the sharp lower bound of $H(x)$ is
\begin{equation}\label{eq:lower}
\underline H(x)
=\max_{J\subseteq[d]}
\left\{
\sum_{T:M(T)\subseteq J}a_T
-\sum_{j\in J}r_j^0
\right\}.
\end{equation}
The sharp upper bound of $H(x)$ is the optimal value of the
fractional packing problem
\begin{equation}\label{eq:upper}
\overline H(x)=
\max\left\{
\sum_{T\subseteq[d]}t_T:
\begin{array}{l}
0\leq t_T\leq a_T
\quad(T\subseteq[d]),\\[2pt]
\displaystyle
\sum_{T:j\in M(T)}t_T\leq r_j^1
\quad(j\in[d])
\end{array}
\right\}.
\end{equation}
Equivalently, the upper bound has the dual
representation
\begin{equation}\label{eq:dual}
\overline H(x)
=\min_{\lambda\in\R_+^d}
\left\{
\sum_{j=1}^d\lambda_jr_j^1
+\sum_{T\subseteq[d]}a_T
\left(1-\sum_{j\in M(T)}\lambda_j\right)_+
\right\},
\end{equation}
where $(u)_+=\max\{u,0\}$. All extrema are attained. In particular, for
each fixed $x\in\R^d$, there exist compatible
full-data laws attaining the lower and upper
bounds, respectively.
\end{theorem}

The lower formula has a direct interpretation.
For any $J\subseteq[d]$, an eligible record whose
missing coordinates are all contained in $J$
can be excluded from the intersection only by
assigning a zero to at least one coordinate in
$J$. At most $\sum_{j\in J}r_j^0$ mass can be
excluded in this way. Consequently, every
compatible full-data law satisfies
\(
H(x)\ge
\sum_{T:M(T)\subseteq J}a_T
-\sum_{j\in J}r_j^0.
\)
Maximizing over $J$ gives a lower bound on
$H(x)$. A capacitated max-flow/min-cut argument
establishes that this bound is sharp, yielding
\eqref{eq:lower}. For the upper
bound, $t_T$ is the eligible pattern-$T$ mass
completed to all ones; this uses one unit of
$r_j^1$ for every $j\in M(T)$, giving
\eqref{eq:upper}. The multipliers $\lambda_j$ in \eqref{eq:dual}
price these missing-coordinate resources.

The lower bound can be evaluated by maximum flow
without enumerating the $2^d$ subsets in
\eqref{eq:lower}. The upper bound is a linear
program with one variable per occurring observation
pattern and $d$ marginal-resource constraints.
Thus, the size of these formulations depends on
the number of occurring patterns, which can be as
large as $2^d$. Section~\ref{sec:lp} of the
Supplementary Appendix provides primal--dual
linear-programming formulations that facilitate
computational implementation and explain why the
lower-bound dual admits an integral optimum,
whereas the upper-bound dual generally does not.

These computational formulations are useful for
bounding probabilities of simultaneous threshold
events, which arise in, for example, multidimensional poverty
\citep{Atkinson2003,DuclosSahnYounger2006},
chronic poverty \citep{Foster2009}, and joint
defaults \citep{Li2000}. When the population
marginals are known but the outcomes are
selectively incompletely observed,
Theorem~\ref{thm:sharp} provides sharp bounds
for these probabilities.

To assess the identifying contribution of incomplete
records, consider the classical pointwise
Fréchet--Hoeffding bounds based on the population
marginals alone:
\begin{equation}\label{eq:classical}
L_{\FH}(x)
=\max\left\{0,\sum_{j=1}^dq_j-(d-1)\right\},
\qquad
U_{\FH}(x)=\min_{j\in[d]}q_j.
\end{equation}
Every admissible completion has these marginals, so
\(
L_{\FH}(x)
\leq\underline H(x)
\leq\overline H(x)
\leq U_{\FH}(x).
\)
Section~\ref{sec:discussion} establishes that
\eqref{eq:lower}--\eqref{eq:upper} reduce exactly
to the classical endpoints when no coordinate is
observed. The following corollary characterizes
when the observed records yield strict improvements.

\begin{corollary}[Identifying quantities and strict tightening]
\label{cor:strict}
At fixed $x$, the sharp endpoints depend on
$(F,\mu)$ only through $(a_T)_{T\subseteq[d]}$
and $(r_j^0,r_j^1)_{j\in[d]}$. Moreover,
\begin{align}
\underline H(x)>L_{\FH}(x)
&\iff \exists J\subseteq[d]:\quad
\sum_{T:M(T)\subseteq J}a_T
-\sum_{j\in J}r_j^0
>L_{\FH}(x),
\label{eq:strictL}\\
\overline H(x)<U_{\FH}(x)
&\iff \exists\lambda\in\R_+^d:\quad
\sum_{j=1}^d\lambda_jr_j^1
+\sum_{T\subseteq[d]}a_T
\left(1-\sum_{j\in M(T)}\lambda_j\right)_+
<U_{\FH}(x).
\label{eq:strictU}
\end{align}
The ordinate $H(x)$ is point-identified if and
only if $\underline H(x)=\overline H(x)$.
\end{corollary}

Simple observable sufficient conditions follow
from Theorem~\ref{thm:sharp}. Taking $\lambda=0$ in
\eqref{eq:dual} yields $\overline H(x)\leq A$,
so $A<U_{\FH}(x)$ ensures strict upper tightening.

For the lower bound, let
$N_0=\sum_{j\in S}(1-D_j)$ count the recorded
zeros. Taking $J=[d]$ in \eqref{eq:lower}
and using $A=\Pp(N_0=0)$ yields the following
inequality, whose derivation is given in the Appendix:
\begin{equation}\label{eq:lowercertificate}
\underline H(x)
\geq\sum_{j=1}^dq_j-(d-1)
+\E[(N_0-1)_+].
\end{equation}
Hence, if $\sum_jq_j\geq d-1$ and
$\Pp(N_0\geq2)>0$, the observed records strictly
improve the classical lower bound. These
sufficient conditions are simpler to check than
the necessary-and-sufficient criteria in
Corollary~\ref{cor:strict}.

\section{Discussion: classical bounds and identifying information}
\label{sec:discussion}

The classical Fr\'echet--Hoeffding bounds provide a
benchmark for assessing the information in incomplete
records. When $\Pp(S=\varnothing)=1$, only
$a_\varnothing=1$ is nonzero, and
$r_j^0=1-q_j$, $r_j^1=q_j$. Equations
\eqref{eq:lower}--\eqref{eq:upper} then recover
$L_{\FH}(x)$ and $U_{\FH}(x)$, with both endpoints
attained. Once outcomes are recorded, the observed-data
law further restricts the set of compatible joint laws
and therefore cannot widen the classical interval.
These restrictions need not yield strict tightening:
for either endpoint, improvement requires ruling out
every classical extremizer, as characterized
in Corollary~\ref{cor:strict}. Section~\ref{sec:refinement}
of the Supplementary Appendix further shows that
recording additional coordinates cannot weaken the
bounds and gives observable certificates of improvement.

The additional identifying information has a particularly
simple form at each fixed threshold vector $x$.
The sharp intersection bounds depend on $(F,\mu)$
only through the eligible pattern masses $(a_T)_T$
and the residual threshold masses $(r_j^0,r_j^1)_j$.
Although the full subdistributions $\mu_T$ are essential
to the observation model and are preserved by the
sharpness construction, their within-threshold detail
does not affect these endpoints. This distinction helps
explain why joint observation is unnecessary for
improvement: recorded threshold probabilities can rule
out classical extremizers even when no record contains
more than one outcome, as illustrated in Section~\ref{sec:supp_mobility}
of the Supplementary Appendix.

The same allocation principle applies in other settings
where different components of a latent outcome vector
are observed for different units. With potential outcomes,
treatment assignment reveals one of $Y(0)$ and $Y(1)$;
externally known population marginals and treatment-specific
subdistributions can tighten bounds on their joint law
without restricting assignment. The construction in
Section~\ref{sec:supp_treatment} of the Supplementary
Appendix reduces the bound on an upward threshold-crossing
probability from $[0,0.50]$ to $[0.30,0.50]$.
This is distinct from bounding average treatment effects
under sample selection \citep{RathnayakeEtAl2026}.
In higher dimensions, split-questionnaire designs create
multiple observation patterns, possibly with nonresponse
\citep{RaghunathanGrizzle1995,AxenfeldEtAl2024}.
The 20-indicator example in Section~\ref{sec:supp_questionnaire}
sharpens $[0,0.84]$ to $[0,0.20]$; with only five occurring
patterns, its upper-bound LP has five variables and 20
resource constraints rather than $5\cdot2^{20}$ binary
cells.

These conclusions concern population identification under
unrestricted dependence between $S$ and $X$; the known
marginals and observed subdistributions must describe the
same target population. Sharpness is generally pointwise,
so a completion attaining a bound at one threshold need
not attain it at another. Section~\ref{sec: duality-Biv Spec} in the Supplementary Appendix
shows that both bounds are uniformly sharp when $d=2$
and $p_\varnothing=0$, but that uniform sharpness can
fail when $p_\varnothing>0$, even if the pointwise
envelopes equal the classical FH bounds. For $d>2$,
the pointwise lower envelope need not itself be a joint
CDF. Beyond intersection probabilities,
Section~\ref{sec:Pt_Idt_Binary_Profile} of the
Supplementary Appendix gives a feasible-support
criterion for point identification of general
binary-profile scores. Sampling uncertainty in
$(F,\mu)$ remains outside the present analysis.

\section{Conclusion}\label{sec:Conclusion}

I establish sharp Fr\'echet--Hoeffding bounds using
known population marginals and selectively incomplete
records, without restricting the observation mechanism.
An exact threshold reduction yields max-flow/min-cut
and fractional-packing representations. Illustrations
involving intergenerational mobility, potential outcomes,
and split questionnaires demonstrate that incomplete
records can substantially sharpen identification, even
without jointly observed outcomes. Accounting for
sampling uncertainty remains for future research.

\appendix
\section*{Appendix: Proofs}
%\section{Compatibility and constrained transport}\label{app:compat}

\begin{proof}[Proof of Proposition~\ref{prop:compact}]
To establish nonemptiness, recall that the true
pattern-specific latent measures are defined by
\(
\pi_T^0(B)=\Pp(X\in B,S=T),
\; B\in\mathcal B(\cX).
\)
By construction,
\(
(\pr_T)_\#\pi_T^0=\mu_T
\;(T\subseteq[d]),
\)
and
\(
\sum_{T\subseteq[d]}(\pr_j)_\#\pi_T^0=F_j
\;(j\in[d]).
\)
Thus $(\pi_T^0)_{T\subseteq[d]}\in\cA(F,\mu)$,
so the admissible family is nonempty.

To establish relative compactness, fix
$(\pi_T)_{T\subseteq[d]}\in\cA(F,\mu)$. The marginal
constraints imply
\(
\sum_{T\subseteq[d]}(\pr_j)_\#\pi_T=F_j,
\; j\in[d].
\)
Since each $\pi_T$ is a nonnegative measure, it follows
that
\(
(\pr_j)_\#\pi_T\leq F_j
\)
as finite measures, for every $T\subseteq[d]$ and
$j\in[d]$.

Fix $\varepsilon>0$. Since each $\cX_j$ is Polish,
the probability measure $F_j$ is tight. Hence, for
every $j\in[d]$, there exists a compact set
$K_j\subseteq\cX_j$ such that
\(
F_j(K_j^c)<\frac{\varepsilon}{d}.
\)
Let $K=\prod_{j=1}^dK_j$, which is compact in
$\cX$. For every admissible family and every
$T\subseteq[d]$,
\begin{align*}
\pi_T(K^c)
\leq
\sum_{j=1}^d
\pi_T\bigl(\pr_j^{-1}(K_j^c)\bigr)
=
\sum_{j=1}^d
(\pr_j)_\#\pi_T(K_j^c)
\leq
\sum_{j=1}^dF_j(K_j^c)
<\varepsilon.
\end{align*}
Thus, for each fixed $T\subseteq[d]$, the collection
\(
\left\{
\pi_T:
(\pi_U)_{U\subseteq[d]}\in\cA(F,\mu)
\right\}
\)
is uniformly tight. Moreover, every such measure
has the fixed total mass $p_T$. By Prokhorov's
theorem, this collection is relatively compact in
the topology of weak convergence of finite measures
on $\cX$.

Since there are finitely many observation patterns,
$\cA(F,\mu)$ is relatively compact in the product
of the weak topologies. Since $\cX$ is Polish,
weak convergence of probability measures is
metrizable. The same holds for finite measures
of fixed mass $p_T$ by normalization when
$p_T>0$; for $p_T=0$, the space is a singleton.
Hence the finite product topology is metrizable,
so sequential closedness suffices.

It remains to establish closedness. Let
$(\pi_T^{(n)})_{T\subseteq[d]}$ be a sequence in
$\cA(F,\mu)$ such that
\(
\pi_T^{(n)}\Rightarrow\pi_T
\;\text{for every }T\subseteq[d],
\)
where $\Rightarrow$ denotes weak convergence of
finite measures. Since the coordinate projections $\pr_T$ and
$\pr_j$ are continuous, weak convergence is
preserved under their pushforward mappings.
Consequently,
\(
(\pr_T)_\#\pi_T
=
\lim_{n\to\infty}(\pr_T)_\#\pi_T^{(n)}
=\mu_T,
\; T\subseteq[d],
\)
where the limit is understood in the weak topology.
Similarly, for every $j\in[d]$,
\(
\sum_{T\subseteq[d]}(\pr_j)_\#\pi_T
=
\lim_{n\to\infty}
\sum_{T\subseteq[d]}(\pr_j)_\#\pi_T^{(n)}
=F_j.
\)
In particular, $\pi_T(\cX)=p_T$ for every $T$.
Thus $(\pi_T)_{T\subseteq[d]}\in\cA(F,\mu)$,
and $\cA(F,\mu)$ is closed.

Combining relative compactness and closedness
establishes that $\cA(F,\mu)$ is weakly compact.
\end{proof}

\begin{proof}[Proof of Proposition~\ref{prop:ot}]
First, each $\pi\in\cA(F,\mu)$ determines a compatible full-data law. On $\cX\times2^{[d]}$, define
$Q_\pi(C)=\sum_T\pi_T(\{z:(z,T)\in C\})$ for $C\in\mathcal B(\cX)\otimes\mathcal P(2^{[d]})$.
Since $\sum_T\pi_T(\cX)=1$, this is a probability measure. Its coordinate mappings $X^\pi(z,T)=z$ and $S^\pi(z,T)=T$ satisfy, for $B\in\mathcal B(\cX_T)$,
$Q_\pi(S^\pi=T,\pr_T(X^\pi)\in B)=\mu_T(B)$ and
$\mathcal L_{Q_\pi}(\pr_j(X^\pi))=F_j$, where $\mathcal L_{Q_\pi}(\pr_j(X^\pi))$ denotes the
probability law of $\pr_j(X^\pi)$ under $Q_\pi$.
Thus the objective in \eqref{eq:bounds} equals $\E_{Q_\pi}[c(X^\pi)]=\sum_T\int c\,\mathrm d\pi_T$.

To obtain the transport coupling, define
$\Phi(z,T)=((T,\pr_T(z)),z)$ and $\gamma_\pi=\Phi_\#Q_\pi$. The map $\Phi$ is measurable and $K$ is Borel, since each $\pr_T$ is continuous and there are finitely many patterns.
For $B\in\mathcal B(\cX_T)$, $\gamma_\pi((\{T\}\times B)\times\cX)=\mu_T(B)$. Hence its $\cO$-marginal is $P_O$; its latent-coordinate marginals are $F_j$, and $\gamma_\pi(K)=1$.
Also, $\int c(z)\,\mathrm d\gamma_\pi(o,z)=\sum_T\int c\,\mathrm d\pi_T$.

Conversely, let $\gamma$ satisfy \eqref{eq:otconstraints} and set
$\pi_T(B)=\gamma((\{T\}\times\cX_T)\times B)$ for $B\in\mathcal B(\cX)$.
Then $\pi_T(\cX)=p_T$ and, for every $C\in\mathcal B(\cX_T)$, concentration on $K$ yields
\begin{align*}
(\pr_T)_\#\pi_T(C)
=\gamma((\{T\}\times\cX_T)\times\pr_T^{-1}(C))
=\gamma((\{T\}\times C)\times\cX)
=P_O(\{T\}\times C)=\mu_T(C).
\end{align*}
Moreover, $\sum_T(\pr_j)_\#\pi_T=\gamma_{X_j}=F_j$, and
$\sum_T\int c\,\mathrm d\pi_T=\int c(z)\,\mathrm d\gamma(o,z)$.
Thus $\pi\in\cA(F,\mu)$, and the two feasible classes have identical sets of objective values, proving the claim.
\end{proof}

%\section{Exact threshold lifting and sharpness}\label{app:lift}
\begin{proof}[Proof of Theorem~\ref{thm:lift}]
For $b\in\{0,1\}^d$ write $B^b=\prod_jB_j^{b_j}$. For $j\in[d]$ and $e\in\{0,1\}$, put
\(
r_j^e=R_j(B_j^e),
\)
where $R_j$ is the residual marginal measure defined in \eqref{eq:residual}. If $\pi\in\cA(F,\mu)$, setting $w_{T,b}=\pi_T(B^b)$ yields the observed-category and marginal equalities in \eqref{eq:W}; hence $w\in\cW(\theta,q)$.

Conversely, fix $w\in\cW(\theta,q)$. The constraints imply, for every $j$ and $e\in\{0,1\}$, $\sum_{T:j\notin T}\sum_{b:b_j=e}w_{T,b}=r_j^e$. Indeed, the sum for $e=1$ is $q_j$ minus observed mass with $D_j=1$, and the sum for $e=0$ follows by subtracting from $m_j$. 
For $r_j^e>0$, define the probability measure
\(
\nu_j^e(C)
=\frac{R_j(C\cap B_j^e)}{r_j^e},
\; C\in\mathcal B(\cX_j).
\)
Similarly, for $\theta_{T,a}>0$, define
\(
\mu_T^a(C)
=\frac{\mu_T(C\cap B_T^a)}{\theta_{T,a}},
\; C\in\mathcal B(\cX_T).
\)
Thus, $\nu_j^e$ and $\mu_T^a$ are the normalized
restrictions of $R_j$ and $\mu_T$ to the
corresponding threshold cells. Whenever $w_{T,b}>0$, both $\theta_{T,b_T}>0$ and $r_j^{b_j}>0$ for every $j\notin T$, so all these distributions are defined.

For each $(T,b)$ with $w_{T,b}>0$, let
$Q_{T,b}$ be the probability law on $\R^d$
obtained by drawing the observed coordinates
jointly from $\mu_T^{b_T}$, the missing
coordinates independently from the measures
$\nu_j^{b_j}$, and reassembling the coordinates
in their original order. By construction,
$Q_{T,b}$ is supported on $B^b$.

Define
\(
\pi_T=\sum_{c:w_{T,c}>0}w_{T,c}Q_{T,c}.
\)
Because the cells $(B^b)_{b\in\{0,1\}^d}$
are pairwise disjoint, for every
$b\in\{0,1\}^d$,
\(
\pi_T(B^b)
=\sum_{c:w_{T,c}>0}
w_{T,c}Q_{T,c}(B^b)
=w_{T,b}.
\)
For any Borel $C\subseteq\R^T$,
\(
(\pr_T)_\#\pi_T(C)=\sum_a\left(\sum_{b:b_T=a}w_{T,b}\right)\mu_T^a(C)
=\sum_a\theta_{T,a}\mu_T^a(C)=\mu_T(C),
\)
with zero-mass categories omitted. Furthermore, the combined marginal contributed by missing-coordinate patterns is
\(
\sum_{T:j\notin T}(\pr_j)_\#\pi_T
=\sum_{e\in\{0,1\}}\left(\sum_{T:j\notin T}\sum_{b:b_j=e}w_{T,b}\right)\nu_j^e
=\sum_e r_j^e\nu_j^e=R_j,
\)
where zero-mass terms are omitted. Adding the recorded-coordinate contribution reproduces $F_j$ by \eqref{eq:residual}. Thus $\pi\in\cA(F,\mu)$ and has the prescribed binary cell masses. 

Under the maintained probability model,
$\cW(\theta,q)$ is nonempty because the true
joint law of $(S,D)$ induces a feasible array
$w^0_{T,b}=\Pp(S=T,D=b)$.
Moreover, $\cW(\theta,q)$ is defined by finitely
many linear equalities and nonnegativity
constraints, and every feasible array satisfies
\(
\sum_{T\subseteq[d]}
\sum_{b\in\{0,1\}^d}w_{T,b}
=\sum_{T\subseteq[d]}
\sum_{a\in\{0,1\}^T}\theta_{T,a}=1.
\)
Thus $\cW(\theta,q)$ is a nonempty compact
polytope. Linear objectives attain their
extrema on this set, and the preceding
construction lifts every optimizer to a
compatible full-data law. Hence both sharp
bounds are attained.

\end{proof}

%\section{Proof of the sharp cut and packing formulas}
%\label{app:sharp}

\begin{proof}[Proof of Theorem~\ref{thm:sharp}]
By Theorem~\ref{thm:lift}, it suffices to
optimize over $\cW(\theta,q)$.

\emph{Lower bound: maximum exclusion.}
The lower bound is obtained by excluding as
much eligible mass as possible from the
intersection. An eligible record with pattern
$T$ can be excluded only by assigning zero
to at least one of its missing indicators.
Let $f_{Tj}\geq0$, for $j\in M(T)$, denote
the eligible pattern-$T$ mass excluded by
assigning zero to missing coordinate $j$.
The total excluded mass is maximized by
\[
\begin{aligned}
V_0=\max_{f\geq0}\quad
\sum_T\sum_{j\in M(T)}f_{Tj}
\;\text{subject to}\;
\sum_{j\in M(T)}f_{Tj}\leq a_T,
T\subseteq[d],\;\text{and}\;
\sum_{T:j\in M(T)}f_{Tj}\leq r_j^0,
j\in[d].
\end{aligned}
\]
The first constraints limit the excluded mass
within each pattern, while the second limit
the total missing-zero mass available for
each coordinate. 

Every feasible binary allocation induces
such an $f$ by assigning each excluded
eligible cell to one of its missing
coordinates with indicator zero.
Conversely, any feasible $f$ can be
realized by partitioning the eligible
pattern-$T$ mass into disjoint portions
of sizes $f_{Tj}$ and assigning zero to
coordinate $j$ on each portion. The remaining
missing indicators can be completed without
violating their marginal constraints, as
shown by Lemma~\ref{lem:partial_completion}
in the Supplementary Appendix.
Consequently,
\(
\underline H(x)=A-V_0.
\)

\emph{Minimum-cut characterization.}
The optimization defining $V_0$ is a
maximum-flow problem. Construct a network
with a source, one node for each pattern
$T$, one node for each coordinate $j$,
and a sink. The source-to-pattern arcs
have capacities $a_T$, the pattern-to-coordinate
arcs have capacity $A+1$ when $j\in M(T)$,
and the coordinate-to-sink arcs have
capacities $r_j^0$.

The max-flow/min-cut theorem
\citep[Theorem~6.3]{AhujaEtAl1993}
equates $V_0$ with the minimum cut capacity.
The interpretation is that a cut identifies
a bottleneck preventing further eligible
mass from being excluded.

To compute this capacity, let $J$ be the
set of coordinate nodes on the source
side of a cut. Since a cut of capacity
$A$ exists, no minimum cut crosses an
arc of capacity $A+1$. Thus a pattern
node $T$ can lie on the source side only
if $M(T)\subseteq J$. For fixed $J$,
placing every such pattern node on the
source side minimizes the cut capacity.
It follows that
\(
V_0=\min_{J\subseteq[d]}
\left\{
\sum_{T:M(T)\nsubseteq J}a_T
+\sum_{j\in J}r_j^0
\right\}.
\)
Subtracting from $A=\sum_Ta_T$
gives \eqref{eq:lower}.

\emph{Upper bound: maximum inclusion.}
For the upper bound, the objective is reversed:
as much eligible mass as possible must be
completed to the all-one event.
Let $t_T$ denote the mass of pattern $T$
assigned to this event. Necessarily,
$t_T\leq a_T$, and each such completion
uses missing-one mass in every coordinate
$j\in M(T)$. Hence
\(
\sum_{T:j\in M(T)}t_T\leq r_j^1,
\; j\in[d].
\)
These are precisely the constraints of
the fractional packing problem
\eqref{eq:upper}.

Conversely, for any feasible $t$, reserve
eligible pattern-$T$ mass $t_T$ and assign
one to all its missing indicators.
The remaining indicators can again be
completed by Lemma~\ref{lem:partial_completion}.
The resulting allocation belongs to
$\cW(\theta,q)$ and has intersection mass
at least $\sum_Tt_T$. Since every feasible
allocation satisfies the packing constraints,
the optimum in \eqref{eq:upper} is the
sharp upper bound.

\emph{Duality and attainment.}
The LP dual of \eqref{eq:upper} assigns
nonnegative multipliers $\lambda_j$ to the
missing-one resource constraints and
$\tau_T$ to the pattern-capacity constraints.
These multipliers can be interpreted as
prices for the resources needed to complete
eligible records to the intersection.
The dual constraints are
\(
\tau_T+\sum_{j\in M(T)}\lambda_j\geq1.
\)
Eliminating $\tau_T$ gives
\(
\tau_T=
\left(1-\sum_{j\in M(T)}\lambda_j\right)_+,
\)
and hence \eqref{eq:dual}.
Strong LP duality gives equality of the
primal and dual optimal values, with
dual attainment.

Finally, Theorem~\ref{thm:lift} lifts the
optimizing binary allocations to compatible
full-data laws. This establishes pointwise
sharpness and attainment of both bounds.
\end{proof}

\begin{proof}[Proof of Corollary~\ref{cor:strict}]
The dependence claim is immediate from
\eqref{eq:lower}--\eqref{eq:dual}.
Since these are attained maxima and minima,
each strict inequality with the classical
endpoint holds exactly when some corresponding
cut $J$ or nonnegative multiplier $\lambda$
improves that endpoint. Equality of the two
sharp endpoints is equivalent to point
identification because the feasible class is
convex and both endpoints are attained.
\end{proof}

\begin{proof}[Proof of the observable certificate
\eqref{eq:lowercertificate}]
Write
$N_0=\sum_{j\in S}(1-D_j)
\leq\sum_{j=1}^d(1-D_j)$.
For each binary vector $D$, the pointwise identity
\(
\prod_{j=1}^dD_j
=
1-\sum_{j=1}^d(1-D_j)
+
\left(\sum_{j=1}^d(1-D_j)-1\right)_+
\)
holds. Since $v\mapsto(v-1)_+$ is increasing for
$v\geq0$, replace the final count with $N_0$,
take expectations, and use $\E[D_j]=q_j$.
This gives \eqref{eq:lowercertificate}.
\end{proof}

% ============================================================
% Supplementary Appendix (within the complete arXiv document)
% ============================================================
\clearpage
\setcounter{page}{1}
\renewcommand{\thepage}{S\arabic{page}}
\setcounter{section}{0}
\setcounter{equation}{0}
\setcounter{theorem}{0}
\renewcommand{\thesection}{S\arabic{section}}
\renewcommand{\thetable}{S\arabic{table}}
\renewcommand{\thefigure}{S\arabic{figure}}
% Distinct PDF destinations for the supplement's sections.
\renewcommand{\theHsection}{supp.\arabic{section}}
\hypersetup{pageanchor=false}

\begin{center}
{\LARGE \bfseries Supplementary Appendix to\\[0.4em]
\emph{Sharp Fr\'echet--Hoeffding Bounds under Selectively Incomplete Observation}\par}
\vspace{1.3em}
{\large Rami V. Tabri\par}
\end{center}
\vspace{1em}

\noindent This supplement collects auxiliary identification results and further illustrations of the bounds in the main manuscript. The proofs of the principal identification results are contained in the appendix of the \emph{main manuscript}, not here. The notation throughout is that of the main manuscript: $S\subseteq[d]$ is the random observation pattern; $T$ denotes a fixed pattern; $\mu_T$ is the observed subdistribution; $F_j$ is the known population marginal; $R_j$ is its residual measure; and $\cA(F,\mu)$ is the class of compatible pattern-specific latent measures. At fixed thresholds, $D_j=\ind\{X_j\leq x_j\}$, $q_j=\Pp(D_j=1)$, $\theta_{T,a}=\Pp(S=T,D_T=a)$, $a_T=\theta_{T,\one_T}$, and $r_j^e=R_j(B_j^e)$. The missing-coordinate set is $M(T)=[d]\setminus T$.

\section{Finite-dimensional duality and bivariate specialization}
\label{sec: duality-Biv Spec}
For a score $g:\{0,1\}^d\to\R$, let $L_g$ and $U_g$ denote its sharp endpoints over $\cW(\theta,q)$. The full linear-programming dual has multipliers for both the observed threshold categories and the marginal probabilities. Eliminating the former leaves an optimization problem in $d$ variables.
\begin{proposition}[Eliminating observed-category multipliers]\label{prop:binarydual}
For $\beta,b\in\R^d$, write $\beta\cdot b=\sum_j\beta_jb_j$. Then
\begin{align}
L_g(\theta,q)&=\max_{\beta\in\R^d}\left\{q\cdot\beta+\sum_{T,a}\theta_{T,a}\min_{b:b_T=a}\bigl[g(b)-\beta\cdot b\bigr]\right\},\\
U_g(\theta,q)&=\min_{\beta\in\R^d}\left\{q\cdot\beta+\sum_{T,a}\theta_{T,a}\max_{b:b_T=a}\bigl[g(b)-\beta\cdot b\bigr]\right\}.
\end{align}
Both extrema are attained.
\end{proposition}
\begin{proof}
Assign unrestricted multipliers $\alpha_{T,a}$ to the observed-category equalities and $\beta_j$ to the marginal equalities in the binary allocation polytope. The lower LP dual has constraints $\alpha_{T,b_T}+\beta\cdot b\leq g(b)$ for each $(T,b)$, and objective $\sum_{T,a}\alpha_{T,a}\theta_{T,a}+q\cdot\beta$. For fixed $\beta$, maximizing over $\alpha$ sets $\alpha_{T,a}=\min_{b:b_T=a}(g(b)-\beta\cdot b)$. The upper dual reverses the inequalities and replaces the inner minima by maxima. The primal feasible polytope is nonempty and compact, so finite-dimensional LP duality gives equality of values and attainment of both dual optima.
\end{proof}

\begin{corollary}[Two-coordinate formulas]\label{cor:bivar}
When $d=2$, put $h=a_{\{1,2\}}$ and abbreviate $a=a_{\{1\}}$, $b=a_{\{2\}}$, $w=p_\varnothing$, $s=p_{\{1\}}$, and $t=p_{\{2\}}$. Set $u=q_1-\Pp(S=\{1,2\},D_1=1)$, $v=q_2-\Pp(S=\{1,2\},D_2=1)$, and $r=s+t+w$. Thus $u$ and $v$ are the threshold-one masses outside the fully observed pattern, and $r$ is the total mass of those patterns. The sharp joint-CDF endpoints are
\begin{align}
\underline H&=h+\max\{0,u+v-r,\ v-s-w+a-b,\ u-t-w-a+b\},\\
\overline H&=h+\min\{u,v,u+v-a-b,a+b+w\}.
\end{align}
\end{corollary}
\begin{proof}
The lower cut formula has four cases $J=\varnothing,\{1\},\{2\},\{1,2\}$, and the full-case eligible mass $h$ belongs to each. The residual identities $r_1^1=u-a$, $r_2^1=v-b$, $r_1^0=t+w-u+a$ and $r_2^0=s+w-v+b$ yield the displayed lower formula. For the upper endpoint $t_{\{1,2\}}=h$ is unconstrained by missing-coordinate resources, leaving the three variables associated with $\varnothing,\{1\},\{2\}$. Its dual attains its minimum at one of $(\lambda_1,\lambda_2)=(0,0),(1,0),(0,1),(1,1)$, resulting in the four terms $u,v,u+v-a-b,a+b+w$.
\end{proof}

\par{\bf Uniform sharpness.}
For $d=2$ and each $\pi\in\cA(F,\mu)$, write $H_\pi(x_1,x_2)=\sum_{T\subseteq[2]}\pi_T((-\infty,x_1]\times(-\infty,x_2])$. 
The pointwise lower bound $\underline H$ is
said to be \emph{uniformly sharp} if there
exists a single $\pi^-\in\cA(F,\mu)$ such that
\[
H_{\pi^-}(x_1,x_2)=\underline H(x_1,x_2)
\quad\text{for all }(x_1,x_2)\in\R^2.
\]
Uniform sharpness of the upper bound
$\overline H$ is defined analogously, with
a possibly different admissible family $\pi^+$.
Both bounds are uniformly sharp if each
has this property. Unlike pointwise sharpness,
uniform sharpness requires the same compatible
full-data law to attain a given endpoint
simultaneously at every threshold pair.

\par{\bf Pointwise versus uniform sharpness of the
bivariate CDF bounds.}
The classical bivariate FH bounds are uniformly sharp: comonotone and countermonotone couplings attain the upper and lower CDF envelopes, respectively, at every threshold pair. Adding observed-data restrictions changes the feasible couplings, so pointwise sharpness need not imply uniform sharpness. The following example shows that this distinction persists even if the pointwise endpoints agree everywhere with the classical FH bounds.

To see this, let $X_1,X_2\in\{0,1,2\}$, with both
population marginals uniform on $\{0,1,2\}$.
Suppose $p_\varnothing=p_{\{1\}}=p_{\{2\}}=1/3$ and $\Pp(X_1=1\mid S=\{1\})=\Pp(X_2=1\mid S=\{2\})=1$.
Corollary~\ref{cor:bivar} gives the classical
bivariate FH endpoints at every threshold pair.
In particular, $\overline H(0,0)=1/3$, $\overline H(1,1)=2/3$, and $\underline H(0,1)=\underline H(1,0)=0$.

Suppose a single compatible joint distribution
attained both displayed upper endpoints.
The marginal probabilities then imply
$\{X_1=0\}=\{X_2=0\}$ and
$\{X_1\leq1\}=\{X_2\leq1\}$ almost surely,
so $X_1=X_2$ almost surely. But the disjoint
events $\{S=\{1\},X_1=1\}$ and
$\{S=\{2\},X_2=1\}$ each have probability
$1/3$, whereas $\Pp(X_1=X_2=1)=1/3$,
a contradiction.

Similarly, simultaneous attainment of the
two displayed lower endpoints would force
the joint distribution to assign probability
$1/3$ to each of $(0,2),(1,1),(2,0)$.
The same observed-pattern restrictions
exclude this distribution. Thus neither
pointwise envelope is uniformly attainable,
even though both are bivariate CDFs.

\begin{proposition}[Uniform sharpness without completely
unobserved records]\label{prop:uniform_bivar}
Suppose $d=2$ and $p_\varnothing=0$. Write $G(x_1,x_2)=\mu_{\{1,2\}}((-\infty,x_1]\times(-\infty,x_2])$, $s_j(x_j)=\mu_{\{j\}}((-\infty,x_j])$, and $\rho_j(x_j)=R_j((-\infty,x_j])$ for $j=1,2$. Here $s_j$ denotes the cumulative mass of the observed singleton-pattern subdistribution, distinct from the residual total mass $m_j$ in the main manuscript.

Then the sharp joint-CDF bounds are
\begin{align*}
\underline H(x_1,x_2)
={}&G(x_1,x_2)
+\bigl(s_1(x_1)+\rho_2(x_2)-p_{\{1\}}\bigr)_+ +\bigl(\rho_1(x_1)+s_2(x_2)-p_{\{2\}}\bigr)_+,
\\
\overline H(x_1,x_2)
={}&G(x_1,x_2)
+\min\{s_1(x_1),\rho_2(x_2)\}+\min\{\rho_1(x_1),s_2(x_2)\}.
\end{align*}
Moreover, there exist compatible joint laws $H^-$ and $H^+$ satisfying $H^-=\underline H$ and $H^+=\overline H$ at every threshold pair. Thus both pointwise envelopes are uniformly sharp and are themselves bivariate CDFs.
\end{proposition}

\begin{proof}
Since $p_\varnothing=0$, every admissible family
satisfies $\pi_\varnothing=0$ and
$\pi_{\{1,2\}}=\mu_{\{1,2\}}$. The observed-data and marginal constraints imply that $\pi_{\{1\}}$ couples $\mu_{\{1\}}$ with $R_2$, whereas $\pi_{\{2\}}$ couples $R_1$ with $\mu_{\{2\}}$.
Conversely, any pair of couplings with these
marginals, together with the fixed measure
$\mu_{\{1,2\}}$, defines an admissible family.
The two couplings can be chosen independently, so the identification problem decomposes into two bivariate FH problems with fixed marginals.

Recall that for finite nonnegative measures
$\alpha,\beta$ on $\R$ with common total mass
$p$, their bivariate FH bounds are
\(
\bigl(\alpha((-\infty,x])
+\beta((-\infty,y])-p\bigr)_+
\)
and
\(
\min\{\alpha((-\infty,x]),
       \beta((-\infty,y])\}.
\)
Both are attained simultaneously at every
$(x,y)$ by the countermonotone and comonotone
couplings, respectively. When $p>0$, these
couplings are obtained from the normalized
marginals using
\[
\bigl(Q_{\alpha/p}(U),Q_{\beta/p}(1-U)\bigr),
\qquad
\bigl(Q_{\alpha/p}(U),Q_{\beta/p}(U)\bigr),
\]
where $U$ is uniform on $(0,1)$ and $Q$
denotes the generalized quantile function.
Multiplying their probability laws by $p$
gives the required subprobability couplings.
When $p=0$, the corresponding coupling is
the zero measure.

Apply these constructions separately to
$(\mu_{\{1\}},R_2)$ and
$(R_1,\mu_{\{2\}})$. Choosing the comonotone
coupling in both patterns and adding the
fixed component $\mu_{\{1,2\}}$ gives a
compatible joint law attaining the displayed
upper bound at every threshold pair.
Choosing the countermonotone coupling in
both patterns similarly attains the lower
bound everywhere. The resulting functions
are joint CDFs, proving uniform sharpness.
\end{proof}

\section{Linear-programming interpretation of the sharp intersection bounds}
\label{sec:lp}

\begin{lemma}[Completion of partial binary allocations]
\label{lem:partial_completion}
Let $\mathcal I$ be a finite index set. For each
$i\in\mathcal I$, specify a mass $m_i\geq 0$,
an observation pattern $T_i\subseteq[d]$, and an
an assignment vector $v_i\in\{0,1,\mathsf{u}\}^d$, where $\mathsf{u}$ denotes an indicator whose binary value has not yet been assigned. Thus $v_i$ encodes a partial assignment of binary indicators, and $v_{ij}\in\{0,1\}$ for all recorded coordinates $j\in T_i$.

Suppose the masses reproduce the observed threshold
categories:
\[
\sum_{\substack{i\in\mathcal I:\,T_i=T\\
                 v_{i,T}=a}}m_i
=\theta_{T,a},
\qquad T\subseteq[d],\quad a\in\{0,1\}^T.
\]
For $j\in[d]$ and $e\in\{0,1\}$, let $c_j^e=\sum_{i:\,j\notin T_i,\ v_{ij}=e}m_i$ be the missing-coordinate mass already assigned value $e$. If $c_j^e\leq r_j^e$ for every $j\in[d]$ and $e\in\{0,1\}$,
then there exist nonnegative masses $z_{i,b}$, indexed by $i\in\mathcal I$ and $b\in\{0,1\}^d$, with $\sum_bz_{i,b}=m_i$ and $z_{i,b}=0$ whenever $b_j\ne v_{ij}$ for an already assigned coordinate $j$. The resulting allocation $w_{T,b}=\sum_{i:\,T_i=T}z_{i,b}$ belongs to $\cW(\theta,q)$.
\end{lemma}

\begin{proof}
For each $j\in[d]$, the total mass on which
coordinate $j$ is missing and remains unassigned is
\[
\begin{aligned}
u_j=\sum_{\substack{i\in\mathcal I:\,j\notin T_i\\v_{ij}=\mathsf{u}}}m_i=r_j^0+r_j^1-c_j^0-c_j^1.
\end{aligned}
\]
The second equality follows because
$r_j^0+r_j^1$ is the total mass on which coordinate
$j$ is missing.

Define
\[
\alpha_j=
\begin{cases}
\dfrac{r_j^1-c_j^1}{u_j},&u_j>0,\\[6pt]
0,&u_j=0.
\end{cases}
\]
The assumed inequalities imply
$0\leq\alpha_j\leq1$. Moreover,
\(
c_j^1+\alpha_j u_j=r_j^1,
\)
and
\(
c_j^0+(1-\alpha_j)u_j=r_j^0.
\)
These identities also hold when $u_j=0$, since
then $c_j^e=r_j^e$ for $e\in\{0,1\}$.

Call $b\in\{0,1\}^d$ a binary completion of $v_i$
if $b_j=v_{ij}$ whenever $v_{ij}\neq\mathsf{u}$.
For each $i\in\mathcal I$ and $b\in\{0,1\}^d$,
set
\[
z_{i,b}
=
m_i\,
\mathbf 1\{b\text{ completes }v_i\}
\prod_{j:\,v_{ij}=\mathsf{u}}
\left[
b_j\alpha_j+(1-b_j)(1-\alpha_j)
\right].
\]
The product allocates $m_i$ among binary completions of $v_i$, using the fraction $\alpha_j$ for each unassigned coordinate $j$ to receive value one. Previously assigned indicators remain unchanged.
Since
$\alpha_j+(1-\alpha_j)=1$, summing over all
completions gives
\(
\sum_{b\in\{0,1\}^d}z_{i,b}=m_i.
\)

Define
\(
w_{T,b}=\sum_{i:\,T_i=T}z_{i,b}.
\)
Because all recorded indicators are preserved,
for every $T\subseteq[d]$ and $a\in\{0,1\}^T$,
\[
\begin{aligned}
\sum_{b:\,b_T=a}w_{T,b}=\sum_{\substack{i:\,T_i=T\\v_{i,T}=a}}\sum_b z_{i,b}=\sum_{\substack{i:\,T_i=T\\v_{i,T}=a}}m_i=\theta_{T,a}.
\end{aligned}
\]

It remains to verify the marginal constraints.
For each $j\in[d]$, the total mass assigned
indicator one when coordinate $j$ is missing
equals
\(
\sum_{T:\,j\notin T}\sum_{b:\,b_j=1}w_{T,b}=c_j^1+\alpha_j u_j=r_j^1.
\)
Adding the contribution from patterns in which
$j$ is observed yields
\[
\begin{aligned}
\sum_{T\subseteq[d]}\sum_{b\in\{0,1\}^d}
b_jw_{T,b}=\sum_{T:\,j\in T}
\sum_{\substack{a\in\{0,1\}^T\\a_j=1}}
\theta_{T,a}
+r_j^1=q_j.
\end{aligned}
\]
Hence the construction satisfies both sets of constraints defining $\cW(\theta,q)$, without changing any prescribed indicator.
\end{proof}

The lower-bound formula in Theorem~\ref{thm:sharp} of the main 
manuscript can also be derived directly from linear
programming duality. Recall that $a_T$ is the eligible
pattern-$T$ mass, $A=\sum_Ta_T$, and $r_j^0$ is the
available missing-zero mass for coordinate $j$.
Let $f_{Tj}\geq0$, for $j\in M(T)$, denote the
eligible pattern-$T$ mass excluded from the intersection
by assigning a zero to missing coordinate $j$.
The maximum excludable mass solves
\[
\begin{aligned}
V_0=\max_{f\geq0}\quad&
\sum_T\sum_{j\in M(T)}f_{Tj}\\
\text{subject to}\quad&
\sum_{j\in M(T)}f_{Tj}\leq a_T,
&&T\subseteq[d],\\
&
\sum_{T:j\in M(T)}f_{Tj}\leq r_j^0,
&&j\in[d].
\end{aligned}
\]
This is a capacitated bipartite flow problem:
coordinate nodes supply missing-zero resources,
while pattern nodes have capacities $a_T$. The proof of Theorem~\ref{thm:sharp} in the main manuscript
establishes that $\underline H=A-V_0$.

Assign nonnegative dual multipliers $u_j$ and
$v_T$ to the coordinate-resource and pattern-capacity
constraints, respectively. The LP dual is
\[
\begin{aligned}
V_0=\min_{u,v\geq0}\quad&
\sum_jr_j^0u_j+\sum_Ta_Tv_T\\
\text{subject to}\quad&
u_j+v_T\geq1,
&&j\in M(T).
\end{aligned}
\]
This is a weighted fractional vertex-cover problem on a bipartite graph. The integral nature of its dual is what allows the lower bound to be expressed as a maximum over subsets $J\subseteq[d]$. Indeed, an optimal dual solution can be chosen with $u_j,v_T\in\{0,1\}$.
For completeness, this integrality property
has an elementary proof. Since all objective
coefficients are nonnegative, the dual variables
may first be truncated to $[0,1]$.
For $s\in(0,1)$, define
\[
\widehat u_j(s)=\ind\{u_j\geq s\},
\qquad
\widehat v_T(s)=\ind\{v_T\geq1-s\}.
\]
If $u_j+v_T\geq1$, then
$\widehat u_j(s)+\widehat v_T(s)\geq1$.
Thus the rounded solution is feasible for every
$s$. Moreover,
\[
\int_0^1\widehat u_j(s)\,\mathrm ds=u_j,
\qquad
\int_0^1\widehat v_T(s)\,\mathrm ds=v_T.
\]
The average objective value of these binary
feasible solutions therefore equals the objective
value of the fractional solution. Hence some
binary feasible solution has no greater cost.

For a binary solution put $J=\{j:u_j=1\}$. Given $J$, the least costly feasible choice is $v_T=\ind\{M(T)\not\subseteq J\}$.
Consequently, strong LP duality gives
\[
V_0=
\min_{J\subseteq[d]}
\left\{
\sum_{j\in J}r_j^0+
\sum_{T:M(T)\not\subseteq J}a_T
\right\}.
\]
Subtracting from $A=\sum_Ta_T$ yields
\[
\underline H
=
\max_{J\subseteq[d]}
\left\{
\sum_{T:M(T)\subseteq J}a_T
-\sum_{j\in J}r_j^0
\right\},
\]
which is the max-flow/min-cut formula~\eqref{eq:lower}
of the main manuscript.

For comparison, the upper-bound packing LP is
\[
\begin{aligned}
\overline H=\max_{t\geq0}\quad&
\sum_Tt_T\\
\text{subject to}\quad&
t_T\leq a_T,
&&T\subseteq[d],\\
&
\sum_{T:j\in M(T)}t_T\leq r_j^1,
&&j\in[d].
\end{aligned}
\]
Its dual assigns nonnegative multipliers
$\lambda_j$ and $\tau_T$ to the resource and
pattern-capacity constraints:
\[
\begin{aligned}
\min_{\lambda,\tau\geq0}\quad&
\sum_j\lambda_jr_j^1+\sum_T\tau_Ta_T\\
\text{subject to}\quad&
\tau_T+\sum_{j\in M(T)}\lambda_j\geq1,
&&T\subseteq[d].
\end{aligned}
\]
Eliminating $\tau_T$ yields
\[
\overline H=
\min_{\lambda\in\R_+^d}
\left\{
\sum_j\lambda_jr_j^1+
\sum_Ta_T
\left(1-\sum_{j\in M(T)}\lambda_j\right)_+
\right\},
\]
as in formula~(\ref{eq:dual}). Unlike the lower-bound
dual, this problem generally has no optimal
binary multiplier solution. The distinction
reflects the different combinatorial structures:
excluding an eligible record requires a zero
in at least one missing coordinate, whereas
including it requires a one in every missing
coordinate.

\section{Information refinements and observable certificates}
\label{sec:refinement}
An observation-pattern refinement reveals an additional coordinate while leaving the latent outcome vector, its population distribution, and all population marginals unchanged. Thus the refined data impose additional restrictions on the same underlying identification problem. To compare the resulting endpoints, write the lower-cut and upper-dual objectives as
\begin{align}
C_J(a,r^0)&=\sum_{U:M(U)\subseteq J}a_U-\sum_{k\in J}r_k^0,\\
\Psi(\lambda;a,r^1)&=\sum_k\lambda_kr_k^1+\sum_Ua_U\left(1-\sum_{k\in M(U)}\lambda_k\right)_+.
\end{align}
\begin{proposition}[Recording an additional coordinate]\label{prop:refine}
Fix a pattern $T\subseteq[d]$ and $j\notin T$, and set $T^+=T\cup\{j\}$. On the same underlying probability space define $\widetilde S=T^+$ when $S=T$, and $\widetilde S=S$ otherwise; only the recorded coordinates change. Put $\xi^e=\Pp(S=T,D_j=e)$ and $\eta=\Pp(S=T,D_T=\one_T,D_j=1)$. The refined eligible masses and residual resources satisfy
\begin{align*}
\widetilde a_T&=0,\qquad \widetilde a_{T^+}=a_{T^+}+\eta,\qquad \widetilde a_U=a_U\quad(U\notin\{T,T^+\}),\\
\widetilde r_j^e&=r_j^e-\xi^e,\qquad \widetilde r_k^e=r_k^e\quad(k\ne j).
\end{align*}
For every $J\subseteq[d]$ and $\lambda\in\R_+^d$,
\[
C_J(\widetilde a,\widetilde r^0)\geq C_J(a,r^0),\qquad
\Psi(\lambda;\widetilde a,\widetilde r^1)\leq\Psi(\lambda;a,r^1).
\]
Consequently $\underline H\leq\underline{\widetilde H}\leq\overline{\widetilde H}\leq\overline H$. A strict inequality at an originally maximizing cut (respectively minimizing multiplier) is sufficient for strict lower (respectively upper) tightening.
\end{proposition}
\begin{proof}
Write $N=M(T^+)$, so $M(T)=N\cup\{j\}$. Direct substitution gives
\[
C_J(\widetilde a,\widetilde r^0)-C_J(a,r^0)
=\xi^0\ind\{j\in J\}+\eta\ind\{N\subseteq J\}-a_T\ind\{N\cup\{j\}\subseteq J\}.
\]
When $j\notin J$, this is either $0$ or $\eta$. When $j\in J$ but $N\nsubseteq J$, it is $\xi^0$. In the remaining case it is $\xi^0+\eta-a_T=\Pp(S=T,D_j=0,D_T\ne\one_T)\geq0$.

For the upper objective define $u=(1-\sum_{k\in N}\lambda_k)_+$ and $v=(1-\lambda_j-\sum_{k\in N}\lambda_k)_+$. Then $0\leq u-v\leq\lambda_j$, and
\[
\Psi(\lambda;\widetilde a,\widetilde r^1)-\Psi(\lambda;a,r^1)
=-\lambda_j\xi^1+\eta u-a_Tv
=-\lambda_j\xi^1+\eta(u-v)-(a_T-\eta)v\leq-\lambda_j(\xi^1-\eta)\leq0.
\]
Maximizing over $J$ and minimizing over $\lambda$ yield the endpoint inequalities. The refined observation law is generated from the same complete outcome vector, ensuring compatibility. Strictness follows by evaluating a refined cut or dual at an optimizer for the original bound.
\end{proof}

For a fixed threshold vector, let $N_0=\sum_{j\in S}(1-D_j)$ count observed threshold failures. For each $k\in[d]$, let $\omega_k=\Pp(k\in S,D_k=1,\ \exists j\in S:\ D_j=0)$ measure the observed mass where coordinate $k$ succeeds but another recorded coordinate fails. These quantities yield simple certificates of strict tightening without solving the full optimization problems.
\begin{proposition}[Observable tightening certificates]\label{prop:cert}
Let $A=\sum_Ta_T$. The sharp bounds satisfy
\[
\underline H(x)\geq\sum_jq_j-(d-1)+\E[(N_0-1)_+],\qquad
\overline H(x)\leq\min\{A,\min_k(q_k-\omega_k)\}.
\]
In particular, if $\sum_jq_j\geq d-1$ and $\Pp(N_0\geq2)>0$, the classical lower FH bound is strictly improved. Also, $A<\min_jq_j$, or $\omega_k>0$ for some $k$ attaining $\min_jq_j$, strictly improves the upper FH bound.
\end{proposition}
\begin{proof}
The eligible mass equals $A=\Pp(N_0=0)$, and $\sum_jr_j^0=d-\sum_jq_j-\E[N_0]$. Taking $J=[d]$ in the sharp lower cut formula yields
\[
\underline H\geq A-\sum_jr_j^0=\sum_jq_j-(d-1)+\E[N_0]-\Pp(N_0\geq1)
=\sum_jq_j-(d-1)+\E[(N_0-1)_+].
\]
Taking $\lambda=0$ in the upper dual gives $\overline H\leq A$. Taking $\lambda=e_k$, the $k$th unit vector, gives $\overline H\leq r_k^1+\sum_{T:k\in T}a_T=q_k-\omega_k$. When $\sum_jq_j\geq d-1$, a positive probability of at least two recorded threshold failures makes the lower certificate strictly larger than the classical FH endpoint. The remaining strictness claims follow by comparing these inequalities with the classical FH endpoints.
\end{proof}

\section{Point identification of general binary-profile scores}
\label{sec:Pt_Idt_Binary_Profile}
The following statements concern point identification of any score $g:\{0,1\}^d\to\R$, not only the intersection indicator. Write the binary allocation constraints as $Gw=v$, where $v=(\theta,q)$. The matrix $G$ has one row for each observed category $(U,a)$, with entry $\ind\{T=U,b_U=a\}$ in column $(T,b)$, and one row for each marginal $j$, with entry $b_j$ in that column. The \emph{feasible support} is
\[
E=\{(T,b):\text{some }w\in\cW(\theta,q)\text{ has }w_{T,b}>0\}.
\]
Let $G_E$ denote the submatrix formed by columns indexed by $E$, and set $c_E=(g(b))_{(T,b)\in E}$. Write $\operatorname{row}(G_E)$ for the \emph{row space} of $G_E$, namely the linear subspace of $\R^E$ spanned by its rows.
\begin{proposition}[Feasible-support criterion]\label{prop:point}
The score $\E[g(D)]$ is point-identified if and only if $c_E\in\operatorname{row}(G_E)$. Equivalently, there exist coefficients $\alpha_{T,a}$ and $\beta_j$ such that
\[
g(b)=\alpha_{T,b_T}+\sum_j\beta_jb_j,\qquad (T,b)\in E.
\]
When these hold, the identified value is $\sum_{T,a}\alpha_{T,a}\theta_{T,a}+\sum_j\beta_jq_j$.
\end{proposition}
\begin{proof}
The row-space representation immediately implies constancy of the score over the feasible polytope. Conversely, if $g$ is point-identified, average for every $e\in E$ one feasible allocation with positive $e$th coordinate. The average $w^\circ$ is strictly positive on $E$ and zero off $E$. For every $z\in\ker(G_E)$, sufficiently small positive and negative perturbations $w_E^\circ\pm\epsilon z$ remain nonnegative and feasible, so constancy of the objective implies $c_E^\top z=0$. Thus $c_E\in\ker(G_E)^\perp=\operatorname{row}(G_E)$.
\end{proof}

The representation in Proposition~\ref{prop:point} gives the identification criterion an observable interpretation. On the feasible support, the score can be written as the sum of an observed-category term $\alpha_{T,b_T}$ and a linear combination $\sum_j\beta_jb_j$ of marginal indicators. Its expectation therefore depends only on the observed probabilities $\theta_{T,a}$ and the known marginal probabilities $q_j$. Conversely, if no such representation exists on $E$, feasible changes in unobserved cell masses alter the score while preserving both sets of identifying constraints.

\begin{corollary}[A full-support characterization]\label{cor:fullsupport}
Let $\mathcal T_+=\{T:p_T>0\}$ and $J=\bigcap_{T\in\mathcal T_+}T$. Suppose some feasible $w^\circ$ is strictly positive in every cell $(T,b)$ with $T\in\mathcal T_+$ and $b\in\{0,1\}^d$. Then $\E[g(D)]$ is point-identified if and only if for some $\beta\in\R^d$ and some $h:\{0,1\}^J\to\R$,
\[
g(b)=h(b_J)+\sum_{j=1}^d\beta_jb_j,\qquad b\in\{0,1\}^d.
\]
If $J\ne[d]$, the $k$-of-$d$ headcount $\Pp(\sum_jD_j\geq k)$ is not point-identified for any $k\in\{1,\ldots,d\}$.
\end{corollary}
\begin{proof}
The support assumption makes $E=\{(T,b):T\in\mathcal T_+,\ b\in\{0,1\}^d\}$. By Proposition~\ref{prop:point}, the score less a linear combination of marginal indicators must depend only on $b_T$ for each occurring pattern $T$. If $j\notin J$, at least one occurring $T$ omits $j$, so that difference cannot depend on $b_j$. Repeating over all $j\notin J$ gives a function $h(b_J)$. The reverse implication holds because $J\subseteq T$ for every occurring pattern, making $h(b_J)$ observed for every record. Finally, if $j\notin J$, the increment in the $k$-of-$d$ score when $b_j$ switches from zero to one equals $\ind\{\sum_{i\ne j}b_i=k-1\}$, which is not constant in $b_{-j}$, contradicting the proposed representation.
\end{proof}

The set $J$ contains exactly those coordinates recorded under every occurring observation pattern. Under the stated full-support condition, the criterion says that only a function of these universally observed indicators, together with additive marginal terms, is point-identified. In particular, the nonlinear $k$-of-$d$ headcounts cannot be identified if some coordinate is sometimes unobserved.

\section{Sharpness constructions for the three-generation illustration}
\label{sec:supp_mobility}
Consider a constructed population with $d=3$, $q_1=q_2=q_3=0.2$, and four equiprobable observation patterns
$\varnothing,\{1\},\{2\},\{3\}$. Under each singleton
pattern $T=\{j\}$, $\Pp(D_j=1\mid S=T)=0.5$. Thus $a_\varnothing=0.25$, $a_{\{j\}}=0.125$, $r_j^1=0.075$, and $r_j^0=0.675$ for $j=1,2,3$. Write $e_j\in\{0,1\}^3$ for the vector with a single one in coordinate $j$, and write $\mathbf 0$ and $\mathbf 1$ for the all-zero and all-one vectors.

\paragraph{Three-generation intersection.}
The fractional-packing constraints for the upper endpoint are
\[
t_\varnothing+\sum_{\ell\in[3]\setminus\{j\}}t_{\{\ell\}}\leq0.075,
\qquad j=1,2,3.
\]
Summing gives
$3t_\varnothing+2\sum_{j=1}^3t_{\{j\}}\leq0.225$.
Because $t_\varnothing\geq0$, the objective satisfies
$t_\varnothing+\sum_jt_{\{j\}}\leq0.1125$.
Equality holds for $t_\varnothing=0$ and
$t_{\{j\}}=0.0375$ for each $j$. For an explicit attaining allocation, under each $S=\{j\}$ assign masses $0.0375$, $0.0875$, and $0.125$ to $D=\mathbf1$, $D=e_j$, and $D=\mathbf0$, respectively. Under $S=\varnothing$, assign its entire mass $0.25$ to $D=\mathbf0$. The marginal
probabilities equal $3(0.0375)+0.0875=0.20$ for each
coordinate and the intersection mass is $0.1125$.

To attain the lower endpoint zero, under $S=\varnothing$
assign mass $0.075$ to each $e_j$ and $0.025$ to $\mathbf0$. Under each $S=\{j\}$, assign mass $0.125$ to $e_j$ and $0.125$ to $\mathbf0$. This reproduces all observed categories
and marginal probabilities, but assigns no mass to
$(1,1,1)$. Theorem~\ref{thm:lift} of the main manuscript lifts
both allocations to the original real-valued model.

\paragraph{At least two bottom-quintile generations.}
When at most one coordinate is recorded,
\[
\ind\{D_1+D_2+D_3\geq2\}
\leq\sum_{j\notin S}D_j.
\]
Consequently,
$\Pp(D_1+D_2+D_3\geq2)\leq\sum_jr_j^1=0.225$.
The lower-endpoint allocation above has no binary vector
with two or more ones, so it also attains zero for this
score. To attain $0.225$, in each singleton pattern
$S=\{j\}$ assign mass $0.0375$ to each of the two
vectors with ones in coordinate $j$ and exactly one other coordinate, mass $0.05$ to $e_j$, and mass $0.125$ to $\mathbf0$. Again assign all empty-pattern mass to $\mathbf0$.
Each singleton pattern has observed success mass
$2(0.0375)+0.05=0.125$; every coordinate has total
success mass $0.125+2(0.0375)=0.20$; and exactly
$6(0.0375)=0.225$ mass has two ones. With only the
marginals $q_j=0.2$, the upper bound is $0.3$,
achieved by placing mass $0.1$ on each of $(1,1,0)$,
$(1,0,1)$, and $(0,1,1)$ and the remainder on
$(0,0,0)$.

\paragraph{Continuous-income realization.}
The allocations are compatible with continuous income
outcomes: conditional on $D_j=1$, take $X_j$ uniform on
$(0,1)$, and conditional on $D_j=0$, take $X_j$ uniform
on $(1,2)$, using the same within-category laws
under each pattern. These choices yield the stipulated
marginals $F_j(1)=0.2$ and observed subdistributions,
which are preserved by the different binary completions.
Thus the calculations are not restricted to a binary-data
model.

\section{Potential-outcome copula bounds under selective treatment assignment}
\label{sec:supp_treatment}

Let $Y(0)$ and $Y(1)$ denote the two potential outcomes,
and let $Z\in\{0,1\}$ denote the realized treatment status.
Initially suppose the realized outcome $Y(Z)$ is observed
for every unit. This is the observation-pattern model of
the main manuscript with $X_1=Y(0)$, $X_2=Y(1)$,
$S=\{1\}$ if $Z=0$ and $S=\{2\}$ if $Z=1$.
Treatment assignment may depend arbitrarily on the two
potential outcomes. Assume their full-population marginal
laws $F_0,F_1$ are known from external information
referring to the same target population. Such knowledge
is an additional identifying input, not a consequence of
observing the treatment groups in a nonrandomized study.

For fixed thresholds $y_0,y_1$, write $q_z=F_z(y_z)$, $p_z=\Pp(Z=z)$, and $a_z=\Pp(Z=z,Y(z)\leq y_z)$ for $z=0,1$.
The observed subdistributions identify $a_0,a_1$,
whereas the known population marginals identify
$q_0,q_1$. Compatibility requires, among other things,
$0\le q_0-a_0\le p_1$ and $0\le q_1-a_1\le p_0$.

\begin{proposition}[Sharp potential-outcome intersection bounds]
\label{prop:supp_treat}
For compatible observed subdistributions and externally
known potential-outcome marginals, the sharp bounds on
$H(y_0,y_1)=\Pp(Y(0)\le y_0,Y(1)\le y_1)$ are
\begin{align}
\underline H(y_0,y_1)
={}&\bigl(a_0+q_1-a_1-p_0\bigr)_+
  +\bigl(q_0-a_0+a_1-p_1\bigr)_+, \label{eq:treatL}\\
\overline H(y_0,y_1)
={}&\min\{a_0,q_1-a_1\}
  +\min\{q_0-a_0,a_1\}. \label{eq:treatU}
\end{align}
Both endpoints are attainable without restricting the
assignment mechanism. For continuous marginal laws these
are also sharp bounds on the potential-outcome copula
at $(q_0,q_1)$.
\end{proposition}
\begin{proof}
Under $Z=0$, the threshold masses for $Y(0)$ and
$Y(1)$ are $a_0$ and $q_1-a_1$, respectively, with
total group mass $p_0$. The bivariate
Fr\'echet--Hoeffding bounds for the group-specific
intersection therefore give
\[
\bigl(a_0+q_1-a_1-p_0\bigr)_+
\leq \Pp(Z=0,Y(0)\le y_0,Y(1)\le y_1)
\leq\min\{a_0,q_1-a_1\}.
\]
Likewise, under $Z=1$ the two threshold masses are
$q_0-a_0$ and $a_1$, with total mass $p_1$, yielding
\[
\bigl(q_0-a_0+a_1-p_1\bigr)_+
\leq \Pp(Z=1,Y(0)\le y_0,Y(1)\le y_1)
\leq\min\{q_0-a_0,a_1\}.
\]
The two group-specific bounds are simultaneously attainable: each group has its own feasible coupling, and the marginal distribution of the missing potential outcome in that group is fixed by subtracting the observed subdistribution in the other group from the known population marginal.
Any compatible coupling of the resulting two marginal
laws within each group preserves the full observed
subdistributions and the total population marginal laws.
Thus summing the group-specific endpoints gives sharp
bounds; equivalently, these specialize Theorems~\ref{thm:lift}
and~\ref{thm:sharp} of the main manuscript.
\end{proof}

\par{\bf Uniform sharpness.}
Since every unit has exactly one observed
potential outcome, $p_\varnothing=0$.
Proposition~\ref{prop:uniform_bivar}
therefore establishes that both bounds
in \eqref{eq:treatL}--\eqref{eq:treatU}
are uniformly sharp. In particular, a
single compatible full-data law attains
each endpoint function simultaneously
at all threshold pairs, even under
unrestricted treatment assignment.

\par{\bf A constructed treatment-selection example.}
Suppose $p_0=p_1=1/2$ and the externally known
potential-outcome marginals satisfy
$F_0(c)=F_1(c)=1/2$ at a common threshold $c$.
Suppose $\Pp(Y(0)\le c\mid Z=0)=\Pp(Y(1)\le c\mid Z=1)=0.8$: in each realized-treatment group, 80 percent of observed outcomes fall at or below $c$.
Hence $a_0=a_1=0.4$ and $q_0=q_1=0.5$. The
proposition yields
\begin{equation}\label{eq:treatExample}
0\leq H(c,c)\leq0.20,
\qquad\text{versus}\qquad
0\leq H(c,c)\leq0.50
\quad\text{from the marginals alone}.
\end{equation}
In particular, the probability of crossing the threshold from below under control to above under treatment is $\Pp(Y(0)\le c,Y(1)>c)=q_0-H(c,c)\in[0.30,0.50]$, compared with the marginal-only interval $[0,0.50]$.
Because $q_0=q_1$, the opposite-direction crossing
probability is equal to this probability. Thus the
marginal threshold shares show no net change, but the
observed pattern-specific information establishes
substantial individual-level movements in both directions.

For an explicit sharpness verification, let
$D_z=\ind\{Y(z)\le c\}$. The lower endpoint
$H(c,c)=0$ is attained by placing mass $0.4$ on
$(D_0,D_1)=(1,0)$ and mass $0.1$ on $(0,1)$
under $Z=0$, and mass $0.4$ on $(0,1)$ and $0.1$
on $(1,0)$ under $Z=1$. The upper endpoint
$H(c,c)=0.2$ is attained under $Z=0$ by placing
masses $(0.1,0.3,0.1)$ on $(1,1),(1,0),(0,0)$,
respectively, and under $Z=1$ by placing these
masses on $(1,1),(0,1),(0,0)$, respectively.
Both allocations give the stipulated observed shares
and population marginal shares. They can be lifted to
identical continuous observed-outcome subdistributions
by using, for both potential outcomes, the uniform law
on $(0,1)$ conditional on $D_z=1$ and the uniform law
on $(1,2)$ conditional on $D_z=0$, with $c=1$.

\paragraph{Interpretation and further outcome selection.}
If assignment is independent of the pair
$(Y(0),Y(1))$, then $a_z=p_zq_z$; the displayed
bounds reduce to the classical bivariate
Fr\'echet--Hoeffding bounds. Selective assignment can
instead make the observed subdistributions informative
about dependence, conditional on the externally known
marginals. If an additional outcome-response indicator
$R$ is present, the observation pattern can be set to
$S=\varnothing$ when $R=0$, and to the singleton
corresponding to the realized treatment when $R=1$.
The general framework of the main manuscript continues
to apply, although the simple formulas
\eqref{eq:treatL}--\eqref{eq:treatU} are specific
to the case without additional nonresponse.

This distinction also separates the present illustration
from Rathnayake, Negi, Bartalotti, and Zhao (2026),
\emph{Difference-in-Differences with Sample Selection}
(arXiv:2411.09221v3). They develop sharp bounds for
average treatment effects under outcome selection in
DiD models, using restrictions such as parallel trends,
potential-selection ignorability, and selection
monotonicity. Here the target is instead a joint
potential-outcome probability, and the identifying input
is externally known population marginals rather than
those assumptions.

\section{High-dimensional split questionnaires and selective nonresponse}
\label{sec:supp_questionnaire}

Split-questionnaire designs administer different groups of questions
to different sampled units, with potential additional item or unit
nonresponse; see Raghunathan and Grizzle (1995) and Axenfeld et al.
(2024), cited in the main manuscript. Here the complete outcome is a
vector of $d=20$ indicators recording whether a household meets each
of 20 economic-security criteria. The parameter $H=\Pp(D_1=\cdots=D_{20}=1)$ is the share of households meeting all criteria, while $1-H$ is the share failing at least one. The full vector is conceptually defined
for every sampled unit, including nonrespondents. Population marginal
probabilities are treated as identified by compatible auxiliary
information, without claiming that such data are available for every
survey application.

\paragraph{Randomized modules with selective unit response.}
Partition $[20]$ into four disjoint modules $G_1,\ldots,G_4$,
each containing five indicators. Let $A\in\{1,\ldots,4\}$ be
an assigned module, uniformly randomized independently of $D$.
A unit either reports every item of its assigned module or reports
none. The resulting realized observation pattern is $S=G_A$ or
$S=\varnothing$. The response decision may depend on the full
latent outcome vector and is not assumed ignorable.

Consider the following constructed population data: each complete
module has probability $p_{G_k}=0.20$, the empty pattern has
$p_\varnothing=0.20$, and conditional on $S=G_k$ exactly one of
the five recorded indicators equals zero, with its position
uniform over $G_k$. Independently known population marginals satisfy
$q_j=\Pp(D_j=1)=0.84$ for every $j$. Thus $a_{G_k}=0$, $a_\varnothing=0.20$, $r_j^1=0.84-0.20(4/5)=0.68$, and $r_j^0=0.16-0.20(1/5)=0.12$.
Theorem~4.1 of the main manuscript yields
\begin{equation}\label{eq:split_random}
[\underline H,\overline H]=[0,0.20],\qquad
[L_{\mathrm{FH}},U_{\mathrm{FH}}]=[0,0.84].
\end{equation}
Indeed, none of the observed complete modules is eligible for the
all-one event, so only the empty pattern contributes to the upper
packing objective. Hence $\overline H\le p_\varnothing=0.20$.
For the lower cut formula, every proper subset $J\subsetneq[20]$
has eligible mass zero, while the full set yields
$0.20-20(0.12)<0$. Thus $\underline H=0$.

Both endpoints in \eqref{eq:split_random} are attained by joint
models that also respect randomized assignment. To see this, for
each module let $K_k$ indicate that all five of its indicators
equal one, with $\Pp(K_k=1)=0.20$. Conditional on $K_k=0$,
set exactly one indicator in $G_k$ to zero, uniformly over the five
positions; conditional on $K_k=1$, set all five to one. Assign
$A$ uniformly and independently of the entire vector $D$, and let
the unit report the assigned module exactly when $K_A=0$.
Then $\Pp(S=G_k)=\tfrac14(0.80)=0.20$, the recorded module
has one uniformly located zero, $\Pp(S=\varnothing)=0.20$,
and each marginal equals $0.20+0.80(4/5)=0.84$.

For the upper endpoint, couple $K_1=\cdots=K_4=K$ for a common
Bernoulli variable of mean $0.20$; then $H=0.20$.
For the lower endpoint, make the four events $\{K_k=1\}$
pairwise disjoint, each of probability $0.20$, and take all four
indicators $K_k$ equal to zero on the remaining probability $0.20$.
Then $H=0$. Both models induce the same full observed module
subdistributions (after fixing common within-category distributions)
and the same known marginals. They differ only in unobserved
cross-module dependence.

\paragraph{A variant with binding residual-resource constraints.}
The next example uses only the unrestricted realized observation-pattern model: unlike the preceding construction, it does not assume independent randomized assignment of modules. It illustrates the packing calculation when residual-resource constraints, rather than only the eligible-mass certificate, determine the upper endpoint. Consider the same five realized patterns, each having mass $0.20$. Given $S=G_k$, all five recorded indicators equal one
with probability $0.80$; otherwise exactly one equals zero, uniformly
over the module. Take the externally known margins to be $q_j=0.50$.
Then $a_\varnothing=0.20$, $a_{G_k}=0.16$, $r_j^1=0.50-0.20(0.96)=0.308$, and $r_j^0=0.492$.
Writing $t_0=t_\varnothing$ and $t_k=t_{G_k}$, the upper packing
constraints for each $j\in G_k$ reduce to
\[
t_0+\sum_{\ell\ne k}t_\ell\le0.308,
\qquad k=1,\ldots,4.
\]
If $u=t_0+\sum_{k=1}^4t_k$ is the objective, summing the four
distinct constraints gives $3u+t_0\le4(0.308)$, and hence
$u\le4(0.308)/3$. Equality is feasible at
$t_0=0$ and $t_k=0.308/3\le0.16$ for all $k$.
The lower cut formula is zero: the only positive eligible-pattern
contributions for proper $J$ require $|J|\ge15$, whose zero-resource
cost $0.492|J|$ already exceeds the total eligible mass $0.84$;
$J=\varnothing$ gives zero. Therefore the sharp bounds are
\begin{equation}\label{eq:split_packing}
[\underline H,\overline H]=
\left[0,\frac{4(0.308)}{3}\right]
=[0,0.410\overline6],\qquad
[L_{\mathrm{FH}},U_{\mathrm{FH}}]=[0,0.50].
\end{equation}

For upper attainment, place $0.308/3$ of each pattern-$G_k$
mass on the all-one vector and set all missing indicators to zero
on the remaining pattern-$G_k$ mass. Set every indicator to zero
under the empty pattern. These assignments give each $j$ the
observed success mass $0.192$ plus missing success mass
$3(0.308/3)=0.308$, hence the stipulated marginal $0.50$.
For lower attainment, place the empty-pattern mass on the
all-zero vector; for each $G_k$, choose one fixed missing
coordinate in $G_{k+1}$ (cyclically) and give it value zero
on the $0.16$ eligible mass. No eligible record can then belong
to the all-one event. For each coordinate there remains at least
$0.60-0.16=0.44$ of as-yet-unassigned mass in the three
nonempty patterns omitting that coordinate, enough to allocate
its required $0.308$ missing success mass. Filling these
coordinates independently at appropriate rates preserves the
observed within-module categories and all margins. Theorem~\ref{thm:lift}
lifts both allocations to complete real-valued outcome laws.

If independent randomized assignment is part of a survey's design, that additional information must be respected when claiming sharpness relative to the richer experiment. The first construction shows that genuine sharp tightening is possible even with randomized assignment; the second isolates the role of the residual-resource constraints under unrestricted realized patterns.

\paragraph{Dimension and observation-pattern support.}
With five occurring patterns, direct binary allocation over their
support would have $5\cdot2^{20}=5{,}242{,}880$ potential
cell variables, compared with five variables and 20
marginal-resource inequalities in the upper packing problem.
The lower endpoint can be obtained by a max-flow calculation.
This is a support-dependent reduction, not a polynomial-time
claim in unrestricted $d$. The constructed examples have whole-module
or unit nonresponse; ordinary item nonresponse produces further
realized subpatterns, which the same identification formulation
accommodates. Neither example relies on an imputation model.
\end{document}